\documentclass[a4paper,fleqn]{cas-dc}

\usepackage{upgreek}
\usepackage{graphicx}
\usepackage{amsmath,amssymb,amsfonts}
\usepackage{amsthm}
\usepackage[numbers]{natbib}

\def\tsc#1{\csdef{#1}{\textsc{\lowercase{#1}}\xspace}}
\tsc{DG}
\tsc{LL}
\tsc{FB}
\tsc{LF}
\tsc{LAG}

\begin{document}
\let\WriteBookmarks\relax
\def\floatpagepagefraction{1}
\def\textpagefraction{.001}

\shorttitle{Ultrashort Two-Pulse Train Generation on a 100 TW Laser Beamline Using a Delay Mask After the Final Focusing Optics}    

\shortauthors{D. Gregocki et~al.}  

\title [mode = title]{Ultrashort Two-Pulse Train Generation on a 100 TW Laser Beamline Using a Delay Mask After the Final Focusing Optics}  


\tnotetext[1]{Correspondence: david.gregocki@ino.cnr.it, luca.labate@ino.cnr.it} 

%
\author[1,2]{David Gregocki}[type=editor,
       orcid=0009-0001-3064-8188
       ]

\cormark[1]

\credit{Conceptualization, Methodology, Software, Validation, Formal analysis, Investigation, Data curation, Writing-original draft preparation, Visualization}

\affiliation[1]{organization={Intense Laser Irradiation Laboratory (ILIL), Consiglio Nazionale delle Ricerche - Istituto Nazionale di Ottica (CNR-INO)},
            addressline={Via Giuseppe Moruzzi 1}, 
            city={Pisa},
            citysep={}, 
            postcode={56124}, 
            country={Italy}}

\affiliation[2]{organization={Department of Physics, University of Pisa},
            addressline={Largo Bruno Pontecorvo 3}, 
            city={Pisa},
            citysep={}, 
            postcode={56127}, 
            country={Italy}}

\author[1]{Luca Labate}[
       orcid=0000-0002-0804-4548
       ]
\cormark[1]
\credit{Conceptualization, Validation, Investigation, Resources, Writing-review and editing, Supervision, Project administration, Funding acquisition}

\author[1]{Federica Baffigi}[
       orcid=0000-0002-3554-7351
       ]
\credit{Validation, Investigation}

\author[1]{Lorenzo Fulgentini}[
       orcid=0000-0002-2737-7233
       ]
\credit{Validation, Investigation}

\author[1,3]{Leonida A. Gizzi}[
       orcid=0000-0001-6572-6492
       ]
\credit{Conceptualization, Validation, Investigation, Resources, Writing-review and editing, Supervision, Project administration, Funding acquisition}

\affiliation[3]{organization={The Extreme Light Infrastructure (ELI-ERIC)},
            addressline={Za Radnicí 835}, 
            city={Dolní Břežany},
            citysep={}, 
            postcode={252 41}, 
            country={Czech Republic}}


\begin{abstract}
Experimental results showing the generation of an ultrashort two-pulse train using a two-section delay mask placed downstream of the off-axis parabola focusing optics are reported. Based on the initial simulation results, a circular fused silica plate with a central aperture and a measured thickness of $570\pm50$ $\upmu$m used in the quantitative delay calculation was designed to enable two distinct transverse portions of the incident laser pulse to be focused, ideally, with equal intensity. This fulfills one of the requirements of the resonant multi-pulse ionization injection (ReMPI) scheme for laser wakefield acceleration. As an intermediary step, a calibration of Gafchromic EBT4 films against laser fluence was established, enabling direct characterization of the unattenuated transverse beam profile. The experiment was carried out at the CNR-INO Intense Laser Irradiation Laboratory (ILIL) using a \mbox{240 TW} laser system operated at 100 TW as part of the ongoing preparation for the first experimental demonstration of ReMPI. The pulse-to-pulse delay introduced by the delay mask was experimentally measured (at a reduced power, as customary for temporal autocorrelation measurements) to be $\simeq 901\,\mathrm{fs}$, in good agreement with the expected value of $\simeq 888\,\mathrm{fs}$. While the pulse-to-pulse delay is not yet optimized for ReMPI operation at practically relevant plasma densities, the experimental results demonstrate the feasibility of the proposed delay-mask concept.
\end{abstract}




\maketitle

\section{Introduction}
\label{Introduction}
The generation of controlled laser pulse trains is a crucial technological step toward advanced laser-driven acceleration concepts. In particular, the pulse-train generation scheme presented here is intended to enable the Resonant Multi-Pulse Ionization Injection (ReMPI) scheme \cite{Umstadter1994,Tomassini2017,Tomassini2018,Tomassini2020}. For the effective driving of the ReMPI, the pulses within the train are required to exhibit ideally identical peak intensity and a well-defined pulse-to-pulse delay matched to the plasma density. Although its first proof-of-concept experiment has yet to be performed, recent simulation studies \cite{Tomassini2025} have demonstrated its potential to produce ultra-low-emittance electron beams with tunable bunch length, both of which are key requirements of the EuPRAXIA \cite{Assmann2020}. 

Several techniques have been proposed for generating trains of ultrashort laser pulses, including spectral pulse shaping with diffraction gratings \cite{Weiner1988}, cascaded Michelson interferometers \cite{Siders1998, Shalloo2016}, birefringent crystal arrays \cite{Dromey2007}, and waveplate-based spectral filtering \cite{Robinson2007, Shalloo2016}. While these approaches provide flexible control over pulse timing and amplitude, they often suffer from limitations such as pulse stretching, significant energy losses, wavelength-dependent pulse properties, increasing alignment complexity, or poor scalability to large-aperture high-power laser systems. 

The presented pulse-train generation scheme, developed at the ILIL of CNR-INO\cite{Gizzi2021}, takes advantage of wavefront division using a thin transparent plate, acting as a delay mask \cite{Vantaggiato2018,Marasciulli2021,Marasciulli2023}, capable of generating pulse trains with controllable delays (according to the relative mask thickness and group velocity) and balanced intensities, making them particularly attractive for ReMPI-driven laser wakefield acceleration. It can be placed in the path of an incoming laser pulse either before or immediately after the final focusing optics, typically an Off-Axis Parabolic (OAP) mirror, when the ultrashort duration of the pulse(s) must be preserved, as is the case for Laser Wakefield Acceleration (LWFA). In both cases, the spatial properties of the beam are directly translated into the temporal and intensity structure of the generated pulse train in the focal region, making an accurate characterization of the transverse laser profile essential, as the spatial fluence distribution directly impacts both pulse shaping and focusing conditions. However, high-power laser facilities typically operate with large-diameter beams and fluence levels that exceed the damage thresholds of conventional diagnostics, making direct measurements challenging. Standard approaches rely on attenuation and beam resizing, which can compromise the fidelity of the measured profile.

To overcome these limitations, radiochromic films, spe-cifically Gafchromic EBT4, are employed in this work as a diagnostic tool capable of recording the transverse fluence distribution without the need for attenuation or beam reduction. This approach is particularly well suited for large-area beams, as the films provide a direct, high-resolution measurement over centimeter-scale regions. The use of radiochromic films in this context is motivated by recent studies based on Gafchromic EBT3 \cite{He21}, which have demonstrated their effectiveness in accurately measuring laser beam profiles. Building on these results, the present work extends the methodology to EBT4 films.

It should be noted that the delay mask experimentally investigated in this work does not yet provide pulse-to-pulse delays optimized for ReMPI operation at practically relevant plasma densities; however, they serve as an effective proof-of-principle demonstration of the concept. While the final system design for the first proof-of-concept ReMPI experiment is expected to require further refinement, the present study reports on preliminary experimental validation carried out at the main beamline of the ILIL laser system at CNR-INO, with a primary focus on material testing and feasibility assessment. Notably, the delay mask is deliberately positioned downstream of the OAP mirror, representing, to the best of our knowledge, a novel experimental implementation of this technique at this location in a high-power LWFA laser beamline. Most importantly, this approach enables avoiding possible surface damage to the optics downstream of the delay mask due to edge diffraction, which is an issue of main concern when using, as it is the case here, high-power lasers, potentially hindering the adoption of pulse train generation methods based on wavefront division.

The present study therefore aims to demonstrate the feasibility of implementing the delay mask downstream of the OAP mirror, providing an experimental basis for its further development and eventual integration into the ReMPI scheme. In order to properly design the delay mask, a detailed characterization of the transverse fluence distribution of the beam is needed; to this purpose, we used a recently proposed technique \cite{He21}, whose actual implementation on our large-diameter laser beam we also discuss. 

\section{Theoretical Layout}
\subsection{Brief Theoretical Description of Focusing by an Off-Axis Parabolic Mirror}
\label{Brief Theoretical Description of Focusing by an Off-Axis Parabolic Mirror}
As the Off-Axis Parabolic (OAP) mirrors have become an essential part of focusing optics in many laser facilities around the world, their ability to focus ultrashort laser pulses is widely studied via numerical simulations. Starting from the geometry of a typical paraboloid, one can define a Cartesian coordinate system $Oxyz$ with an origin situated at the point of intersection with the paraboloid’s axis of symmetry and its focal length $f_p$ at $\mathbf{x}_\mathbf{p}=\left(0,0,z_f=f_p\right)$. The paraboloid, also referred to as the parent paraboloid, can then be expressed as $z=(x^2+y^2)/4f_p$. 

\begin{figure}
    \centering
    \includegraphics[width=0.9\columnwidth]{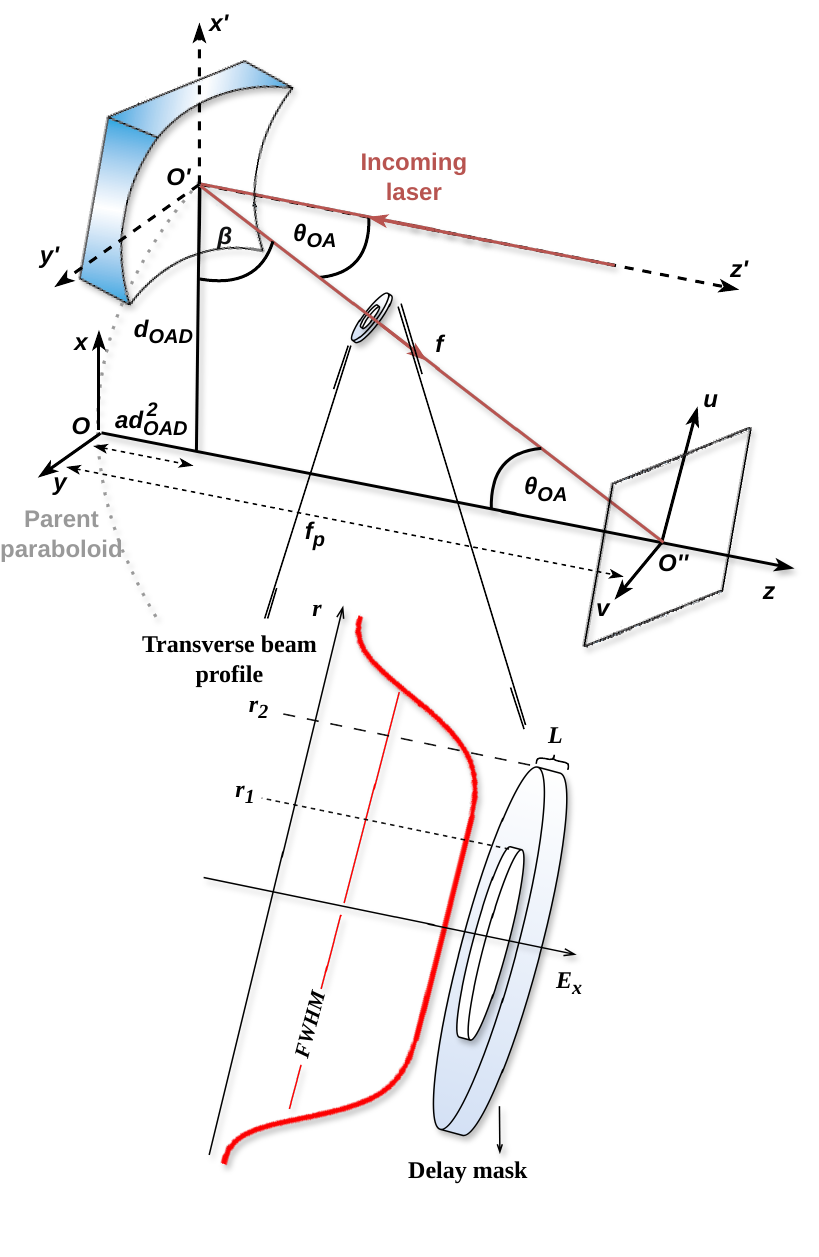}
    \caption{\label{Fig:Figure_1}{Schematic of the coordinate systems employed in the derivation of diffraction integrals for laser pulses focused by an OAP mirror. The right panel shows a close-up of the holed delay mask, which generates a two-pulse train, along with the incident pulse featuring a super-Gaussian transverse profile.}}
\end{figure}

Ultimately, an OAP mirror, with a surface $S_{\mathrm{OAP}}$, can be defined as a section of the parent paraboloid with the origin $O^\prime$ in a new coordinate system $O^\prime x^\prime y^\prime z^\prime$ positioned at the off-axis distance $d_{\mathrm{OAD}}$ from $O$ crossing the paraboloid while taking into account a condition $(x-d_{\mathrm{OAD}})^2+y^2\le\left(\frac{D}{2}\right)^2$, where $D$ is the diameter of the OAP. As can be seen from \autoref{Fig:Figure_1}, several expressions describing the OAP mirror in the defined geometry can be derived as a result. For example,
\begin{align}
    &\tan{(\theta_{OA})}=\frac{d_{\mathrm{OAD}}}{f_p-ad_{\mathrm{OAD}}^2}\quad\text{for}\quad a=\frac{1}{4f_p}, \\
   &f=\sqrt{\left(f_p-ad_{\mathrm{OAD}}^2\right)^2+d_{\mathrm{OAD}}^2}=\frac{d_{\mathrm{OAD}}}{\sin{(\theta_{OA})}}, \nonumber
\end{align}
where $f$ is the (apparent) focal length, $\theta_{\mathrm{OA}}$ is off-axis angle. 

Let’s now consider a laser pulse impinging on the above-described OAP mirror without aberrations propagating in the negative $z^\prime$ direction in the $O^\prime x^\prime y^\prime z^\prime$ system with linear polarization and an electric field without a longitudinal component. In addition, consider an amplitude $A\left(x,y\right)$ characteristic of a laser beam with a super-Gaussian spatial profile
\begin{equation}
    A\left(x,y\right)=A_0\exp{\left(-\frac{1}{2}\left[\left(\frac{x-d_{\mathrm{OAD}}}{\sigma_x}\right)^2+\left(\frac{y}{\sigma_y}\right)^2\right]^{n_{\mathrm{Gauss}}}\right)}.
\end{equation}
Therefore, the electric field of the incident beam already expressed in the $Oxyz$ system can be written as
\begin{equation}
    \mathbf{E}_{\mathrm{Inc}} (\mathbf{x})=A\left(x,y\right)\left(\cos{(\delta)\widehat{\mathbf{e}_\mathbf{x}}}+\sin{(\delta)\widehat{\mathbf{e}_\mathbf{y}}\ }\right)e^{ikp(\mathbf{x})},
\end{equation}
where $\widehat{\mathbf{e}_\mathbf{x}}$, $\widehat{\mathbf{e}_\mathbf{y}}$ are base vectors. Here, the optical path $p\left(\mathbf{x}\right)$ between the point $\left(x,y,ad_{\mathrm{OAD}}^2+z^\prime\right)$ on the reference plane and the point $\left(x,y,a\left(x^2+y^2\right)\right)$ on the OAP surface $S_{\mathrm{OAP}}$ was introduced, while angle $\delta$ accounts for different polarization directions.
Given the system geometry, depicted in \autoref{Fig:Figure_1}, and presented conditions, the diffraction integrals based on the full Stratton-Chu vector theory \cite{Stratton1939} and previous, recent studies \cite{Labate2016,Labate2018} are given as
\begin{align}
    E_j^{\mathrm {Re}}(\mathbf{x_p}, t)=& -\frac{1}{\lambda}  \int_{\text{{S$_{\mathrm{OAP}}$}}}  A(x,y) \times   \nonumber \\
    & \Big[   \mathfrak{Im}\left\{g^{(E_j)}\right \} \cos{(kv-\omega t)} \nonumber \\ 
    &+  \mathfrak{Re}\left\{g^{(E_j)}\right \}  \sin{(kv-\omega t)}  \Big]  \mathrm{d}x \, \mathrm{d}y  , \nonumber \\ 
    E_j^{\mathrm {Im}}(\mathbf{x_p}, t)=& -\frac{1}{\lambda}  \int_{\text{{S$_{\mathrm{OAP}}$}}}  A(x,y) \times \label{Eq:Diffraction_Integrals} \\ 
    & \Big[   \mathfrak{Im}\left\{g^{(E_j)}\right \} \sin{(kv-\omega t)} \nonumber \\ 
    &-  \mathfrak{Re}\left\{g^{(E_j)}\right \}  \cos{(kv-\omega t)}  \Big]  \mathrm{d}x \, \mathrm{d}y  , \nonumber  
\end{align}
and are capable of describing the electric fields of a pulse focused by an OAP mirror. Index $j$ denotes $x$, $z$, $y$ axis, $k$ is the angular wavenumber, \mbox{$v\equiv v\left(x,x_p\right)\equiv p\left(x\right)+u\left(x,x_p\right)$,} while $u\equiv u\left(x,x_p\right)=\left|u\right|=\left|x-x_p\right|$ and $\lambda$ is the wavelength. The terms $g^{\left(E_j\right)}$ can be written as
\begin{align}
    g^{(E_x)}=\frac{1}{u}&\cos{\delta}-\left ( 1-\frac{1}{iku} \right )\frac{1}{u^2} \nonumber \\
    &\times \left ( \frac{x}{2f_p}\cos{\delta}+\frac{y}{2f_p}\sin{\delta} \right )\left ( x-x_p \right ), \nonumber \\
    g^{(E_y)}=\frac{1}{u}&\sin{\delta}-\left ( 1-\frac{1}{iku} \right )\frac{1}{u^2}  \\
    &\times \left ( \frac{x}{2f_p}\cos{\delta}+\frac{y}{2f_p}\sin{\delta} \right )\left ( y-y_p \right ), \nonumber \\
    g^{(E_z)}=\frac{1}{u}&\left ( \frac{x}{2f_p}\cos{\delta}+\frac{y}{2f_p}\sin{\delta} \right )-\left ( 1-\frac{1}{iku} \right )\frac{1}{u^2} \nonumber \\
    &\times \left ( \frac{x}{2f_p}\cos{\delta}+\frac{y}{2f_p}\sin{\delta} \right )\left ( z-z_p \right ) \nonumber.
\end{align}
As a last step, the rotation transformation of the coordinate system $Oxyz$ to $O^\prime xyz$ through $\theta_{\mathrm{OA}}$ around the $y$-axis is necessary to retrieve longitudinal and transverse field components with respect to the focused beam path. From \autoref{Fig:Figure_1}, this path represents the direction along $O^\prime O^{\prime\prime}$.

\subsection{Theoretical Framework of the Delay Mask}

The proposed pulse-train generation scheme employs the delay mask inserted into the beam path, consisting of annular sections with different thicknesses $L$. In the simplest implementation considered here, the mask features a central aperture, i.e., one portion of the incident laser beam passes through the hole without interaction, while the surrounding annular portion propagates through the material. Due to the refractive index difference between air and carefully selected material, the two portions of the beam experience different optical path lengths. This results in a controlled temporal delay $\Delta t$ between the corresponding pulse components upon recombination in the far field. The expected pulse-to-pulse delay is calculated in the usual way using
\begin{equation}
\label{Eq:Pulse_Delay}
    \Delta t=L \left(\frac{1}{v_g}-\frac{1}{c}\right )
\end{equation}
Here $v_g=c/n_g=c/{(n-\lambda\frac{\mathrm dn}{\mathrm d\lambda})} $ where $v_g$ is the group velocity of the laser pulse for a given material, $c$ is the speed of light, $\frac{\mathrm dn}{\mathrm d\lambda}$ is the chromatic dispersion, and $n_g$ is the group refractive index, all terms being calculated at the central wavelength. It follows that the generation of more than two pulses in a train requires a delay mask composed of annular sections with varying thicknesses, with a fixed increment between sections to ensure equal pulse-to-pulse delays. 

In theory, if a laser pulse has a super-Gaussian transverse profile while only its flat-top region impinges on a mask with sections of equal area, i.e., the $N$-th radius of the $N$ section mask fulfills
\begin{equation}
\label{Eq:Mask_Radii}
    r_N=\sqrt N r_1,
\end{equation}
where $r_1$ is the radius of the first, internal section, then each section will carry the same energy, given as a fraction of the energy of the initial pulse and $N$. In addition, under the flat-top beam condition and two-section delay mask configuration, it can be easily verified that
\begin{equation}
\label{Eq:Fraction_Radius_Energy}
    \frac{E_1}{E}=F^2\quad\wedge\quad\frac{A_1}{A_2}=\frac{F^2}{1-F^2} \quad\wedge\quad F=\frac{r_1}{r_2},
\end{equation}
where $E_1, E_2$ are the laser energies within the first and the second mask section of a given areas $A_1=\pi r_1^2$ and \mbox{$A_2=\pi (r_2^2-r_1^2)$,} while $E$ is the energy of the entire laser beam with an area of $A=\pi r_2^2$.  

In practice, however, transverse profiles of laser pulses are not ideally flat, and together with the present tails as sketched in \autoref{Fig:Figure_1} and potential diffractions from the mask edges, individual radii obtained by \autoref{Eq:Mask_Radii} must be adjusted. For this reason, numerical simulations that can retrieve intensity distribution in the focal plane of OAP mirror, e.g., based on the theoretical framework presented in \autoref{Brief Theoretical Description of Focusing by an Off-Axis Parabolic Mirror}, are often required. 

\subsection{Basic Optical Properties of the Delay Mask}
\label{Basic Optical Properties of the Delay Mask}
In this study, an uncoated ultraviolet-grade fused silica plate (Solid Optics CW-0.5-JGS1) with a nominal thickness of \mbox{$L=500\pm100$ $\upmu$m} was selected for the fabrication of the delay mask. The actual measured thickness, determined using a vernier caliper, was \mbox{$L_{\mathrm m}\simeq570\pm50$ $\upmu$m}, well within the manufacturer-stated variation of the nominal value. The value of $L_{\mathrm m}$ will be considered from now on.

Using the Malitson 3-term Sellmeier formula for refractive index $n$ of fused silica at room temperature \cite{Malitson1965}
\begin{align}
\label{Eq:Sellmeier_Equation}
    n^2=1&+\frac{0.6961663\lambda^2}{\lambda^2-0.0684043^2}+\frac{0.4079426\lambda^2}{\lambda^2-0.1162414^2} \\
    &+\frac{0.8974794\lambda^2}{\lambda^2-9.896161^2}, \nonumber
\end{align}
it is possible to numerically obtain $\frac{\mathrm dn}{\mathrm d\lambda}$ and $n_g$. Particularly, at $\lambda=800$ nm, corresponding to the central wavelength of the ILIL laser system, $v_g=204343574.4$ m/s.

\subsection{Applications of the Delay Mask}
As mentioned in \autoref{Introduction}, the main application of the delay mask is its implementation in the ReMPI scheme, where it serves as an optical element for pulse-train generation. One of the requirements for the ReMPI scheme is pulse-to-pulse delay matched to the plasma period, which enables resonant wakefield excitation. In practical application, since the time delay introduced by the delay mask is fixed by its thickness, the plasma density may be adjusted to tune the plasma period to match the imposed delay, thereby allowing resonant wakefield excitation while providing some tolerance to deviations in the fabricated mask thickness. Beyond the ReMPI scheme, the delay mask concept could be potentially extended to practical studies of counter-propagating relativistic laser pulses interacting in a sub-wavelength thin solid-density foil \cite{Jiang2025}, the generation of high-flux angularly uniform protons from multiple laser interactions \cite{Jiang2024}, as well as in multiple-pulse excitation of atoms and molecules \cite{Temkin1993} or microwave communication \cite{Liu1996}.

\section{Methods}
Before presenting the results, we summarize in the following subsections some of the methods used for the mask design, including  the characterization of the laser beam intensity profile via calibrated Gafchromic films.
\subsection{EBT4 Film Response Calibration}
\label{EBT4 Film Response Calibration}
In order to determine the appropriate dimensions of the delay mask, knowledge of the transverse laser profile at the position where the mask will be placed is required. To this purpose, we employed the method recently reported in \cite{He21}, which allows the fluence profile to be gathered via calibrated Gafchromic films.
We used in our case EBT4 films, whose preliminary calibration is briefly reported in this Subsection. The calibration was performed using the ILIL main laser (featuring a pulse duration of $\tau<25$ fs, a central wavelength of $800\,\mathrm{nm}$, and a bandwidth of \mbox{$\sim80\,\mathrm{nm}$}). All measurements, including those reported outside this section, were performed in a single-shot configuration.  

\begin{figure}
    \centering
    \includegraphics[width=\columnwidth]{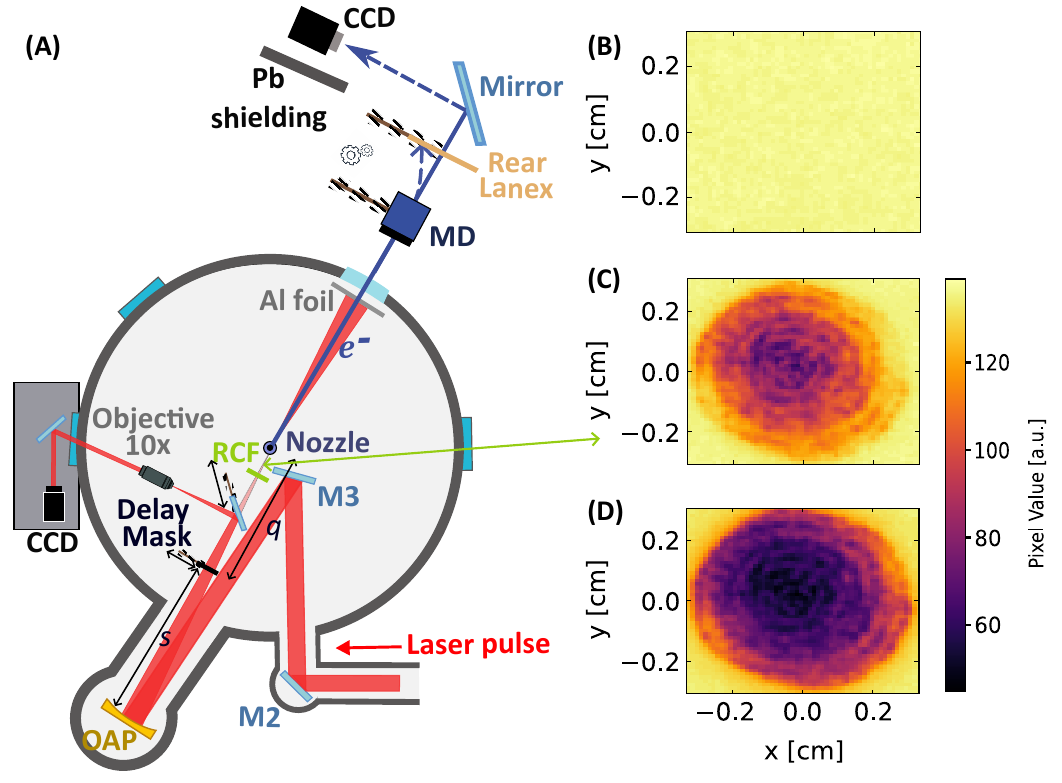}
    \caption{\label{Fig:Figure_2}{(A) Schematic of the experimental setup highlighting the placement of RCF used for the transverse fluence distribution calibration before positioning the delay mask downstream of the OAP mirror, where $s\simeq57$ cm and $q\simeq45$ cm}. During the calibration, the mirror positioned near the focus was removed, allowing RCF irradiation. Transporting mirrors is described with the letter M. On the right is an illustrative example of the EBT4 RCF response to (B) a $0.3~\mathrm{mJ}$, (C) a $1~\mathrm{mJ}$, and (D) a $2~\mathrm{mJ}$ laser beam.}
\end{figure}

The experimental setup, illustrated in \autoref{Fig:Figure_2}, consisted of an OAP mirror with a parent and apparent focal length of \mbox{$f_p\simeq1.016~\mathrm{m}$} and \mbox{$f\simeq1.022~\mathrm{m}$}, respectively and off-axis angle $\theta_{\mathrm{OA}}\simeq8.6^{\circ}$. For the calibration, square $2\times2~\mathrm{cm}$ EBT4 samples were positioned at a distance of $\sim6~\mathrm{cm}$ from the OAP focus and irradiated with different laser energies.

\begin{figure} 
    \centering 
    \includegraphics[width=\columnwidth]{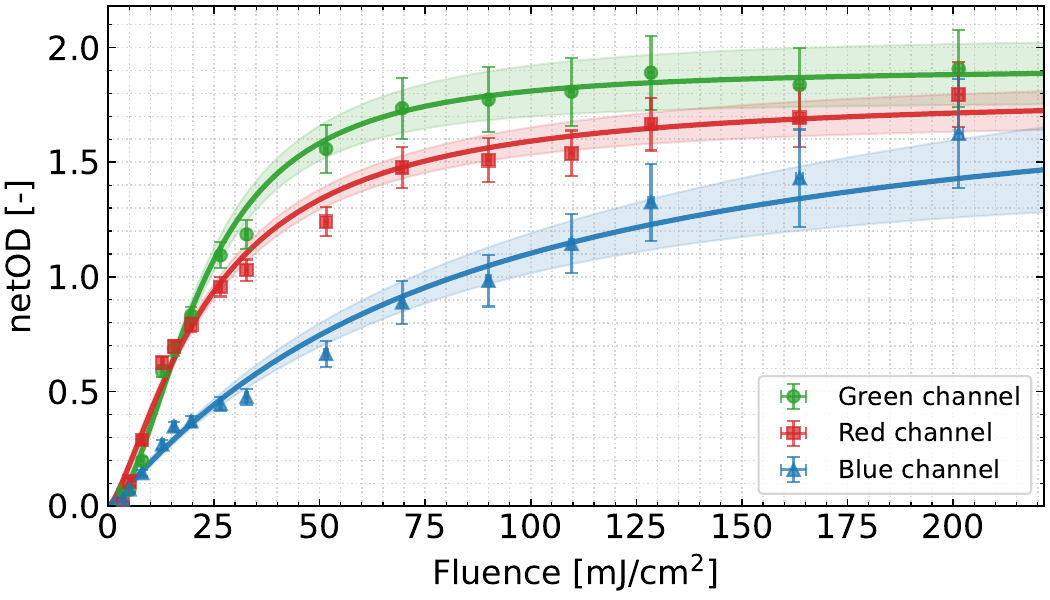} \caption{\label{Fig:Figure_3}{Response curves of netOD to different laser fluences in RGB channels.}} 
\end{figure}

For each channel, the film response was quantified as net optical density (netOD) via mean background-fog subtraction \cite{Gueli2015}:
\begin{equation}
\label{Eq:NetOD_Formula}
    \mathrm{netOD}=\log \left(\frac{\overline{\mathrm{PV}}_\mathrm{Unexp}-\overline{\mathrm{PV}}_\mathrm{Black}}{\mathrm{PV}_\mathrm{Exp}-\overline{\mathrm{PV}}_\mathrm{Black}}\right),
\end{equation}
where PV denotes the pixel value in each RGB channel. The resulting netOD–fluence response curves are shown in \autoref{Fig:Figure_3}, from which it can be seen that the blue channel provides the largest dynamic range.

Each RGB response was fitted using a three-parameter Hill–Langmuir saturation model \cite{Hill1910,Langmuir1918},
\begin{equation}
\label{Eq:Hill_Fit}
    \mathrm{netOD}(\Phi) = \frac{y_0\,\Phi^{n_{\phi}}}{\Phi^{n_{\phi}} + k_{\phi}^{n_{\phi}}},
\end{equation}
where $y_0$, $k_{\phi}$, and $n_{\phi}$ represent the saturation netOD, half-saturation fluence, and Hill coefficient, respectively. Its monotonicity allows  the following inverse relation:
\begin{equation}
\label{Eq:Hill_Inverse}
    \Phi = k_{\phi}\left(\frac{\mathrm{netOD}}{y_0 - \mathrm{netOD}}\right)^{1/n_{\phi}},
\end{equation}
used to fit the fluence against the optical density. The fit of the fluence as a function of netOD was performed channel-by-channel (although the blue channel was ultimately used) via the Levenberg–Marquardt nonlinear least-squares algorithm \cite{Levenberg1944}, with netOD uncertainties used as absolute weights. The results with goodness-of-fit values are given in \autoref{Tab:Calibration_Parameters}.

The netOD uncertainty was obtained from the pixel-value standard deviation within each sampled region and propagated through \autoref{Eq:NetOD_Formula}, while fluence uncertainty was determined from the measured pulse energy and beam area. The $2\sigma$ confidence bands shown in \autoref{Fig:Figure_3} were calculated by full covariance propagation:
\begin{equation}
\label{Eq:ConfidenceBand}
    \sigma_{\mathrm{netOD}}^{2}(\Phi) = \mathbf{J}(\Phi)\,\mathbf{C}\,\mathbf{J}^{\top}(\Phi),
\end{equation}
where $\mathbf{J}$ is the Jacobian of partial derivatives of \autoref{Eq:Hill_Fit} with respect to $(y_0,k_{\phi},n_{\phi})$. This calibration enables measurement of transverse laser-fluence distributions from spatially resolved netOD measurements with a simple implementation and without beam attenuation or downsizing.

\subsection{Designing and Positioning of the Delay Mask}

Following the calibration of the EBT4 films against laser fluence, the dimensions of the delay mask and its individual sections can be determined at the position where the mask is to be inserted into the beam path. In addition, spatial cleaning of the beam using hard aperture is desirable to maximize the uniform, flat-top region of the transverse profile, allowing the subsequent calculations to be performed primarily within this region.

To minimize the risk of laser-induced damage to the transport and focusing optics, the hard aperture was positioned downstream of the OAP mirror rather than upstream in the transport line. The RCF, together with the hard aperture, was placed $s\simeq57~\mathrm{cm}$ after the OAP at an incidence angle of \mbox{$\sim20^\circ$} with respect to the beam propagation axis, see \autoref{Fig:Figure_2}. In this regard, the hard aperture was designed with an elliptical geometry such that its projection onto the plane perpendicular to the laser propagation remained circular, letting through $\sim70\%$ of the impinging laser energy. This aperture was fabricated from polycarbonate material, see \autoref{Fig:Figure_4}, using a CO$_2$ cutter and featured elliptical radii of $b_1\simeq1.921\pm0.005~\mathrm{cm}$ and $b_2\simeq1.802\pm0.005~\mathrm{cm}$, where the uncertainty represents the manufacturer's stated tolerance.
\begin{table}
    \centering
    \begin{tabular}{lcccc}
    \hline
    Channel & $k_{\phi}$ &  $y_0$ & $n_{\phi}$ & $R^2$ \\
    \hline
    Red   & $23.6\pm1.3$  & $1.80\pm0.06$ & $1.42\pm0.07$ & 0.987 \\
    Green & $21.9\pm1.1$ & $1.91\pm0.07$ & $1.91\pm0.09$ & 0.995 \\
    Blue  & $77.1\pm19.4$ & $1.94\pm0.26$ & $1.08\pm0.08$ & 0.978 \\
    \hline
    \end{tabular}
    \caption{Retrieved parameters of \autoref{Eq:Hill_Inverse}, together with the goodness of fit parameter $R^2$.}\label{Tab:Calibration_Parameters}
\end{table}

\subsection{Numerical Simulations and Delay-Mask Optimization}

The primary objective of the simulations carried out for this study was to determine the optimal geometry of the delay mask annular sections, specifically their corresponding areas, such that each section contributed equally to the peak intensity in the focal plane. To do so, the diffraction integrals presented in \autoref{Eq:Diffraction_Integrals} were evaluated using a C++ code developed at ILIL and already benchmarked and employed in previous studies \cite{Labate2016, Labate2018}. The code implements multidimensional integration based on an adaptive cubature algorithm \cite{Berntsen_ACMTMS1991}. The code allows the integration boundaries to be easily defined to simulate the propagation of different annular portions of the beam downstream of the OAP.

\section{Numerical and Experimental results}
\subsection{Transverse Fluence Distribution under High-Power Laser Configuration}

To reproduce the actual experimental geometry of the delay mask, the EBT4 radiochromic film was mounted directly on the rear side of the hard aperture relative to the incident laser beam. The RCF was irradiated with a single laser shot under the high-power laser configuration corresponding to the $\sim 100~\mathrm{TW}$ operational regime. Under these conditions, the laser delivered $\sim2.2~\mathrm{J}$ per pulse before spatial filtering, conservatively accounting for transmission losses introduced by the beam transport optics and, predominantly, by the final compressor stage of the chirped pulse amplification \cite{Strickland1985} system. Assuming a homogeneous transverse fluence distribution, the theoretical fluence right after the hard aperture was estimated to be approximately \mbox{$\Phi_\mathrm{Theo}=138~\mathrm{mJ/cm^2}$.}
\begin{figure}
    \centering
    \includegraphics[width=\columnwidth]{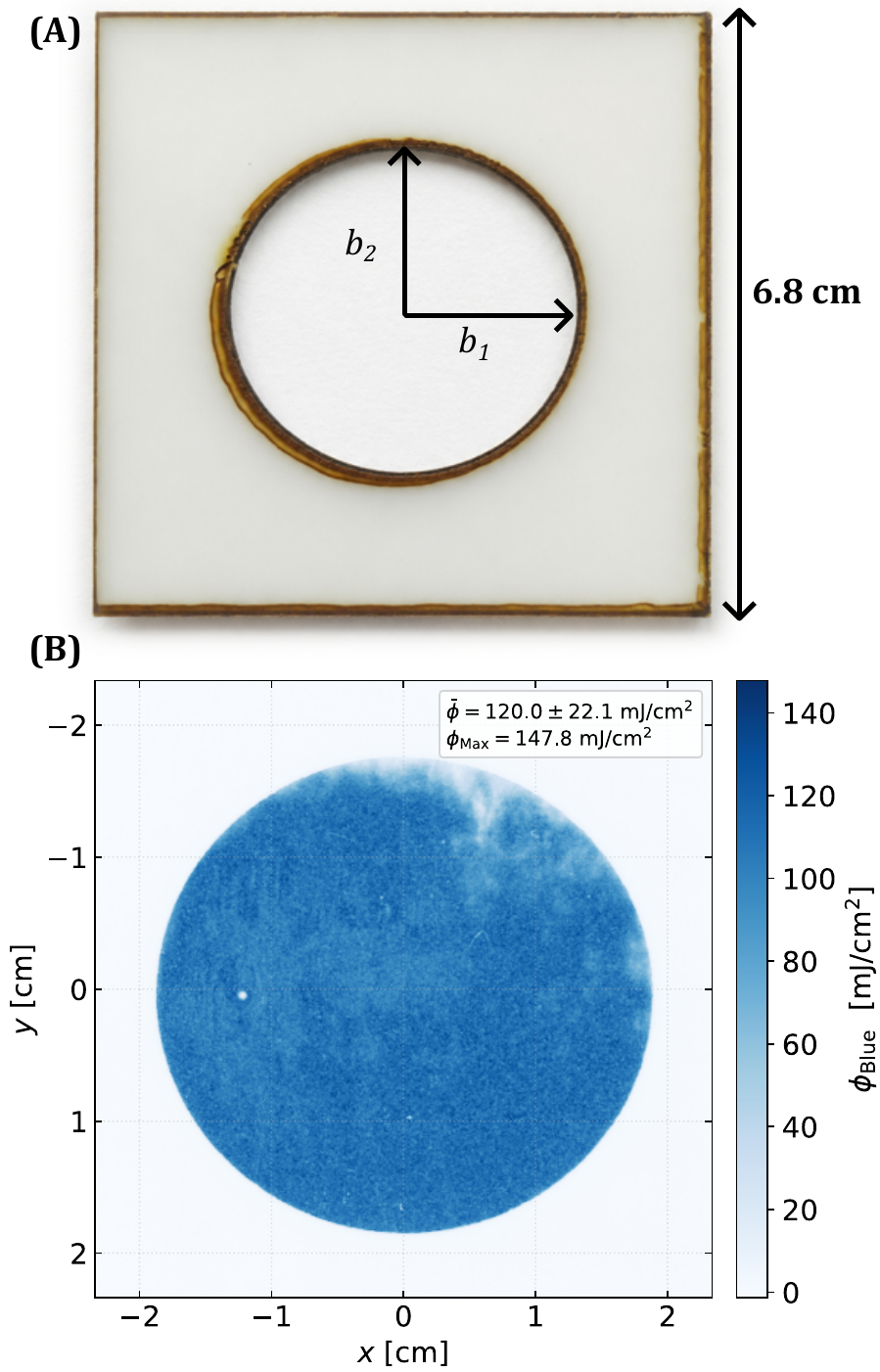}
    \caption{\label{Fig:Figure_4}{(A) Hard aperture used before the RCF for mapping the transverse fluence distribution, and later for the delay mask. (B) Transverse fluence distribution captured by irradiating EBT4 RCF placed behind the hard aperture. To provide a rough estimate of the fluence across the irradiated area of the EBT4 film, the mean $\bar{\Phi}$ and maximum $\Phi_{\mathrm{Max}}$ fluences were calculated as the mean and maximum of the pixel-wise fluence values (retrieved by applying the calibration function to each pixel's netOD). The results are shown in the inset.}}
\end{figure}

After the irradiation, the RCF was scanned, and the measured netOD values were converted pixel-by-pixel into fluence values using \autoref{Eq:Hill_Inverse}, together with the calibration parameters for the blue channel given in \autoref{Tab:Calibration_Parameters}, which was selected for its superior dynamic range. The calibrated transverse fluence distribution of the laser beam under high-power operating conditions is presented in \autoref{Fig:Figure_4}. As can be seen, the retrieved mean fluence \mbox{$\bar{\Phi}\simeq120.0\pm22.1$ mJ/cm$^2$,} where the uncertainty includes the propagated RCF calibration uncertainty and signal-to-noise ratio of the measured RCF response. Both the coordinates and fluence distribution were corrected for RCF tilt and projected onto the plane perpendicular to laser propagation.

\subsection{Dimensions of the Delay Mask}

\begin{figure}
    \centering
    \includegraphics[width=\columnwidth]{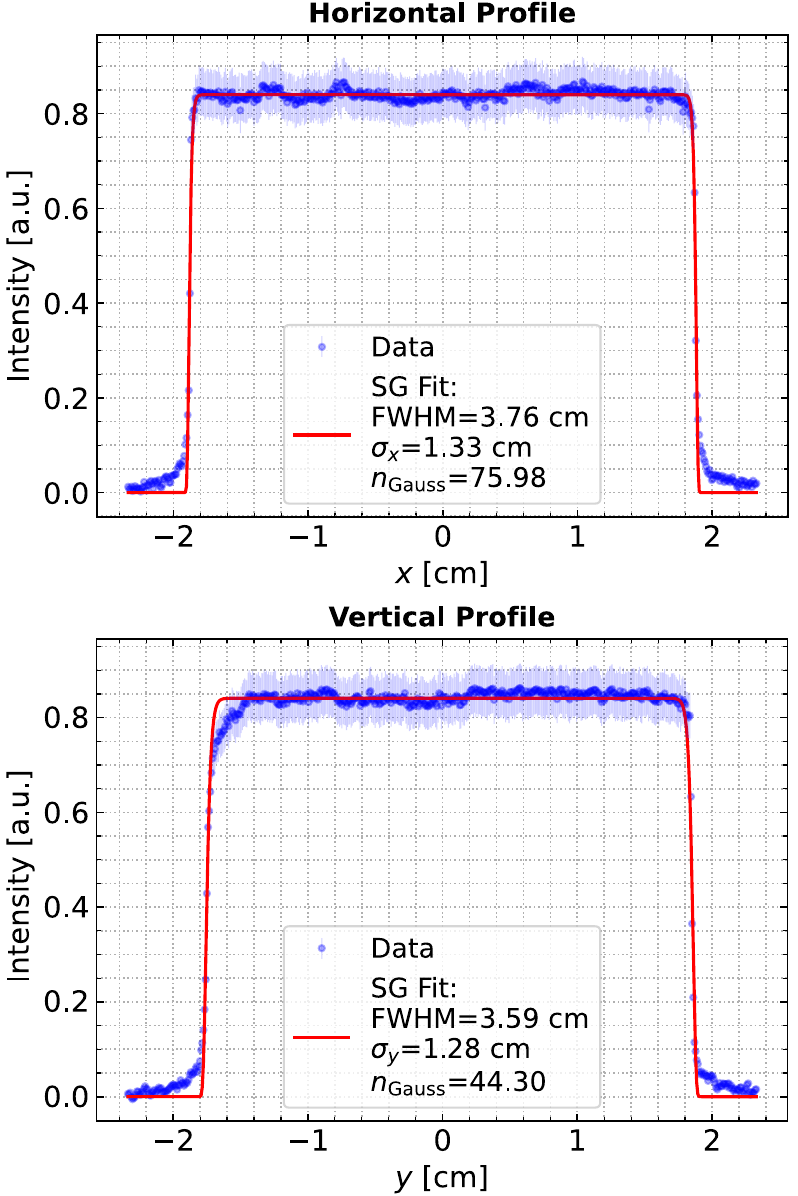}
    \caption{\label{Fig:Figure_5}{1D scans of the intensity profile for an impinging laser pulse in the RCF fitted with a super-Gaussian function in the horizontal ($x$-axis) and vertical ($y$-axis) directions.}}
\end{figure}

\begin{figure*}
    \centering
    \includegraphics[width=\textwidth]{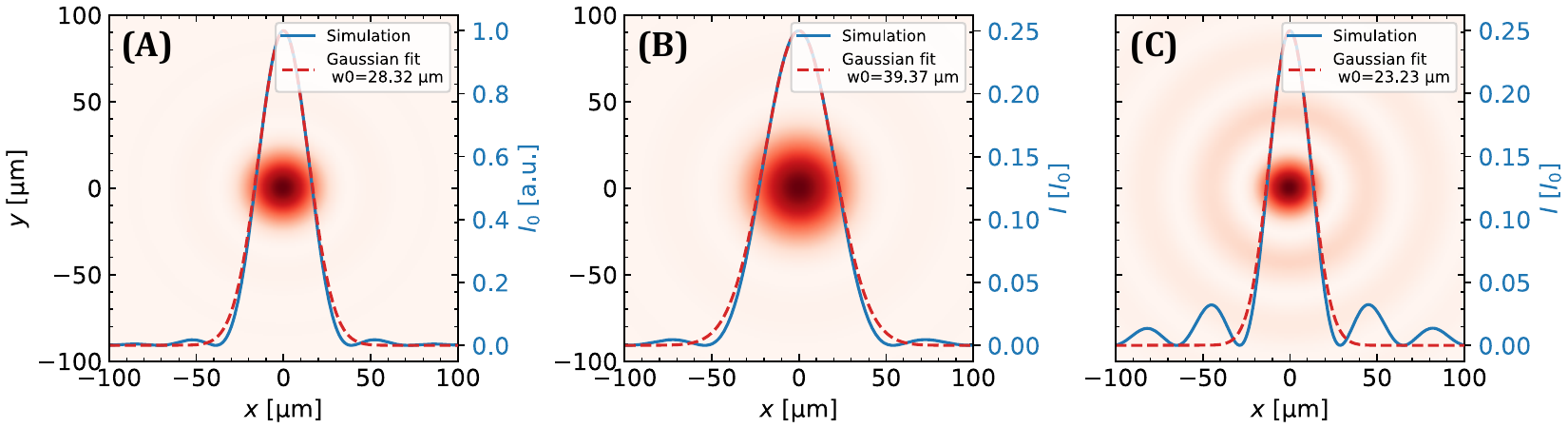}
    \caption{\label{Fig:Figure_6}{Simulated intensity distributions of the laser beam at the focus of the OAP mirror for three configurations: (A) the full laser beam without delay mask interaction, providing a reference, (B) central beam component passing through the mask aperture, (C) the outer annular component interacting with the mask. Solid blue lines represent the 1D intensity profiles along the horizontal direction, with corresponding dashed red lines showing the Gaussian fits.}}
\end{figure*}

Once the transverse fluence distribution of the laser beam was experimentally retrieved, horizontal and vertical lineouts were fitted using a super-Gaussian function, as shown in \autoref{Fig:Figure_5}, to extract the corresponding FWHM, standard deviations $\sigma_x$ and $\sigma_y$, and super-Gaussian order $n_{\mathrm{Gauss}}$. These parameters were subsequently used as inputs for numerical simulations of the laser intensity distribution in the focal plane of the OAP mirror, together with the OAP parameters presented in \autoref{EBT4 Film Response Calibration}
The resulting focal-plane intensity distributions produced by the individual mask sections, optimized to yield identical peak intensities, are presented in \autoref{Fig:Figure_6}. The corresponding inner radii are $r_{1,\mathrm{v}}\simeq9.020\pm0.005~\mathrm{mm}$ and \mbox{$r_{1,\mathrm{h}}\simeq9.431\pm0.005~\mathrm{mm}$,} where the uncertainty  reflects the numerical rounding associated with the simulations.

\subsection{Performance of the Delay Mask}

In order to experimentally test the performance of the final delay mask geometry, the focal intensity distributions produced by the individual mask sections that matched the simulation results were measured in a low-power setting.

\begin{figure}
    \centering
    \includegraphics[width=0.7\columnwidth]{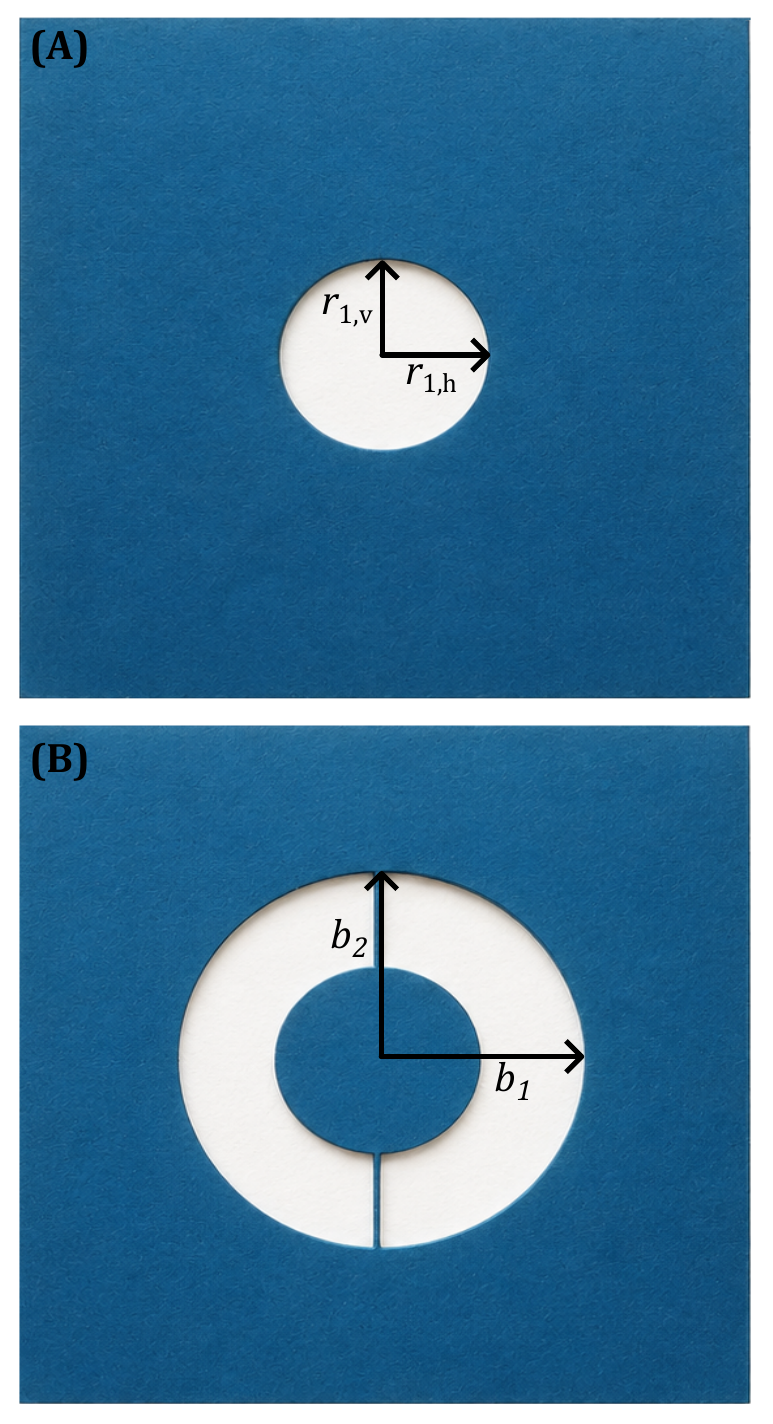}
    \caption{\label{Fig:Figure_7}{Fabricated hard-paper masks used for low-power testing measurements of the delay mask performance. (A) Mask blocking the outer region of the beam while transmitting the central region $A_1$. (B) Mask blocking the central region while transmitting the outer region $A_2$. The geometries of both masks correspond to the dimensions of the final delay mask design used in the simulations}}
\end{figure}

Instead of using the final fused silica delay mask, which does not allow individual annular sections to be selectively blocked, a set of non-transparent spatial masks was fabricated with geometries identical to those of the actual delay mask. The fabricated masks are shown in \autoref{Fig:Figure_7}. Each mask was designed to block either the central region $A_1$ or the outer region $A_2$ of the beam independently.

\begin{figure*} 
    \centering
    \includegraphics[width=\textwidth]{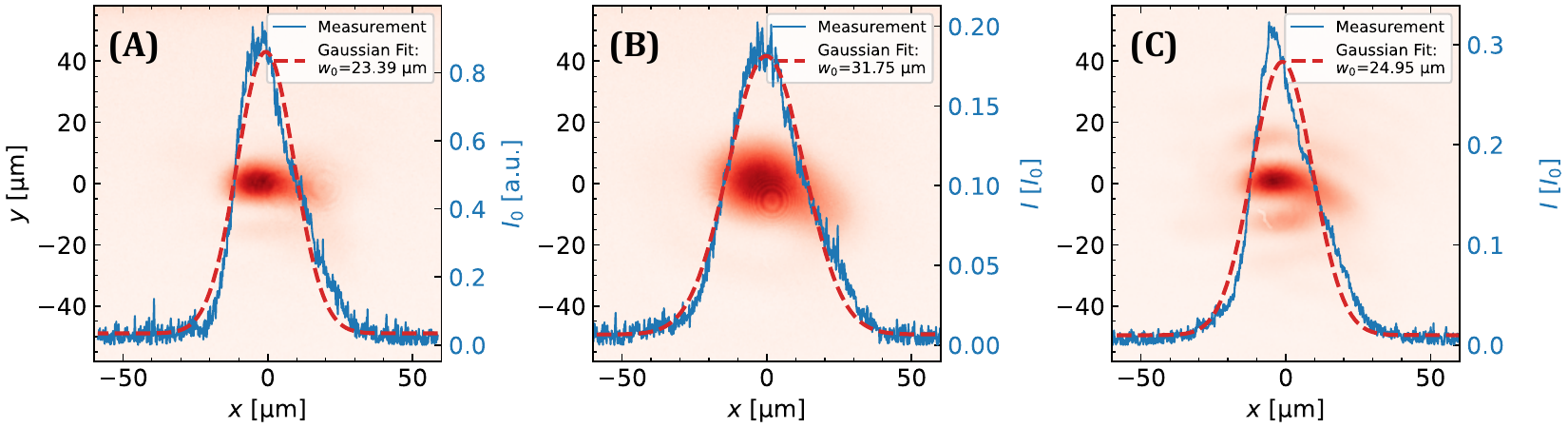}
    \caption{\label{Fig:Figure_8}{Measured intensity distributions of the laser beam at the focus of the OAP mirror for three configurations: (A) the full laser beam without mask interaction, providing a reference, (B) central beam component passing through the mask aperture, (C) the outer annular component interacting with the mask. Solid blue lines represent the 1D intensity profiles along the horizontal direction, with corresponding dashed red lines showing the Gaussian fits.}}
\end{figure*}

By inserting the corresponding paper mask into the beam path, it was possible to isolate the contribution of each delay-mask section and measure the resulting focal-plane intensity distribution individually. The experimentally retrieved focal spots are presented in \autoref{Fig:Figure_8} together with the corresponding fitted one-dimensional horizontal intensity profiles.
A comparison between the simulated and experimentally measured focal-plane distributions shows that the laser waist values closely reproduce those predicted by the simulations.

\begin{figure*}
    \centering
    \includegraphics[width=\textwidth]{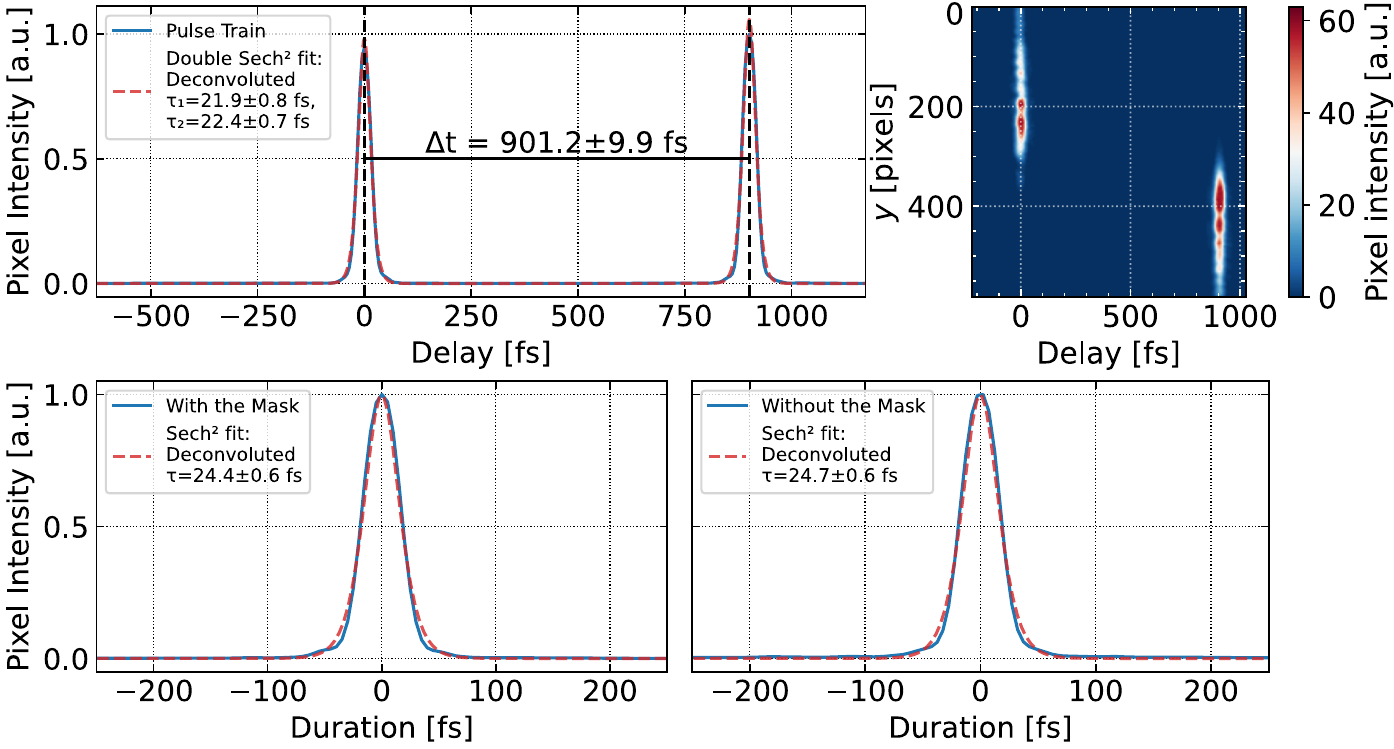}
    \caption{\label{Fig:Figure_9}{Top row: Pulse-to-pulse signal trace retrieved from a single-shot second-order autocorrelator by exploiting its second dimension normally used to improve the signal-to-noise ratio, together with the raw autocorrelator image. The total trace obtained in a single-shot manner is composed of two autocorrelator traces obtained simultaneously. We note that for display purposes, the delay axis is restricted to the relevant region, while the full temporal window available is 4 ps.} Bottom row: Retrieved pulse-duration profiles from autocorrelator traces for the case in which the laser pulse propagates through the delay mask and the reference case without the mask.}
\end{figure*}

The expected pulse-to-pulse delay given by \autoref{Eq:Pulse_Delay} is $\Delta t_\mathrm{Theo}= 888.0\pm78$ fs, considering $L_{\mathrm m}$. Using the single-shot $2^{\mathrm{nd}}$-order autocorrelator (BONSAI, Amplitude Technologies), the pulse-to-pulse delay introduced by the fused silica plate described in \autoref{Basic Optical Properties of the Delay Mask} and placed perpendicular to the beam path was experimentally measured as $\Delta t\simeq901.2\pm9.9~\mathrm{fs}$, using a pulse amplified, for this specific purpose, up to an energy of $\sim 20\,\mathrm{mJ}$, as customary for the autocorrelator measurements. The measured delay, obtained from the trace shown in the top row of \autoref{Fig:Figure_9}, is in good agreement with the expected value $\Delta t_{\mathrm{Theo}}$, with a deviation of \mbox{less than $2\%$.} We note that in the ReMPI scheme, the required pulse-to-pulse delay is determined by the plasma period, which can be adjusted; therefore, this ($\sim2\%$) discrepancy is acceptable.

$\Delta t$ was extracted from a nonlinear least-squares fit of the pulse-train trace to a sum of two sech$^2$ functions, each with its temporal center ($t_1$, $t_2$) and duration ($\tau_1$, $\tau_2$), sharing a common baseline offset. The delay was calculated as the separation of the two fitted peak centers, $\Delta t=t_2-t_1$, while the temporal axis was calibrated via the manufacturer-provided calibration. Because the two peaks in the top panel of \autoref{Fig:Figure_9} are fitted using a single global function with a shared baseline, their slightly different peak intensities cause the fit to marginally overshoot the measured maxima. The $\pm9.9$ fs uncertainty was obtained from the standard deviation of the two-trace peak positions extracted from 1D profiles along the temporal axis at different vertical positions of the raw autocorrelator image, depicted in \autoref{Fig:Figure_9}, accounting for noise, peak-finding uncertainty, and signal fluctuations.
 
We note that the trace presented in the top row of \autoref{Fig:Figure_9} does not have the immediately recognizable form of a conventional second-order autocorrelation. To obtain such a trace in a single-shot manner, the second dimension of the second-order autocorrelation, which is normally used to improve the signal-to-noise (S/N) ratio, was exploited. \autoref{Fig:Figure_9} shows on the top right a raw image from the autocorrelator. Specifically, we used the direction of the autocorrelator perpendicular to the time-encoding direction to get two simultaneous autocorrelation traces from the two regions of the beam and retrieved the pulse separation exploiting the translation invariance along this direction (it is worth observing that we assessed to use a small portion, in both directions, of the raw image from the autocorrelator, where this sort of post-analysis is reliable). 

The pulse widths for the configurations with and without the delay mask, depicted in the bottom row of \autoref{Fig:Figure_9}, were measured assuming a $\mathrm{sech}^2$ temporal profile for the input pulse, and the pulse durations were then obtained using the convolution factor of 1.5426. As can be seen, no significant elongation is observed after propagation through the mask at the power level used for autocorrelator measurements, which is consistent with the expected dispersion and nonlinear effects. In addition, we note that, although the pulse temporal characterisation was performed in a single-shot manner, the delay mask, with a thickness relevant to the ReMPI scheme, is expected to withstand, like other optical elements in the laser chain, at least the $10~{\mathrm{Hz}}$ repetition rate typical of most 100 TW-class lasers used for LWFA.

\section{Discussion}
The results presented in this study highlight several considerations relevant to the design of the delay mask. One minor consideration is the difference between \mbox{$\Phi_\mathrm{Theo}$} and the experimentally retrieved mean fluence \mbox{$\bar{\Phi}$.} The theoretical value is based on the assumption of a uniform transverse fluence distribution, whereas the measured profile exhibits spatial non-uniformities. As shown in \autoref{Fig:Figure_4}, this discrepancy can be attributed to the less uniform region of transverse fluence visible in the upper-right corner. Nevertheless, $\bar{\Phi}$ is consistent with $\Phi_\mathrm{Theo}$ within $\sigma$.

The different focal spot sizes observed in both simulations and experiments (\autoref{Fig:Figure_6} and \autoref{Fig:Figure_8}) originate primarily from wavefront aberrations, which are present in the latter independently from the use of the mask. Nevertheless, the measured focal spots do exhibit the expected trend: the central portion produces a larger focal spot than the outer annular portion, consistent with the simulation prediction.

A notable difference in the peak intensities of the two annular portions is also observed in the experimentally measured focal spots; see \autoref{Fig:Figure_8}. This difference can be attributed, at least in part, to the different laser power levels used for the fluence mapping and focal-spot measurements. This represents a general challenge in ultrashort, high-power laser systems, where diagnostics compatible with operation at the highest power levels are often difficult to implement. In the present case, the mask dimensions were determined from fluence measurements performed at approximately \mbox{100 TW,} whereas the focal spot measurements were acquired at substantially lower laser power, using attenuated $12$ mJ laser pulses, to remain below the damage threshold of the CCD. Since the transverse intensity profile and wavefront are not necessarily invariant with operating conditions and may also change following optimization of the laser chain, this difference can affect both the optimal mask dimensions and the resulting balance between the two pulse intensities. This highlights the importance of performing accurate transverse beam characterization at the power level of the final experiment for an optimal design of the delay mask.

To quantify the spatial alignment of the sub-beams, additional Center-Of-Mass (COM) calculations were conducted using measurements presented in \autoref{Fig:Figure_8}. The focal spots associated with the central and outer annular portions exhibit only minor vertical and horizontal COM shifts relative to the reference beam, of less than $0.08~\upmu\text{m}$ and $0.19~\upmu\text{m}$, respectively. These sub-$\upmu\text{m}$ deviations are negligible relative to the spot size and do not affect the spatial overlap at focus.

Additionally, numerical simulations of the propagation of short pulses in nonlinear media were performed using the Python library \texttt{pychi} \cite{Voumard2023}, considering a central wavelength of 800 nm, fused-silica mask thicknesses of 100 and 570~$\upmu$m, a pulse duration of 25 fs, and a fluence of 0.12~J/cm$^2$. The nonlinear response was described by a Kerr coefficient of $n_2 = 2.5\times10^{-20}$~m$^2$/W, while linear dispersion was modelled using \autoref{Eq:Sellmeier_Equation}. Pulse propagation was simulated using the self-phase modulation model implemented in \texttt{pychi}. For a mask thickness of $100~\upmu$m, which is the most relevant thickness for the intended applications, the simulations yield a $B$-integral of \mbox{$B<0.9$ rad,} suggesting that nonlinear phase accumulation remains sufficiently small, thus supporting the practical use of this approach for high-power operation. In addition, we notice that the simulations yield a negligible pulse broadening in the $100~\upmu$m case, with the pulse duration downstream of the mask being $\simeq 25.6~\mathrm{fs}$. On the other hand, for the $570~\upmu$m thickness case, a significantly higher value of the $B$-integral ($B\simeq5$ rad) can be found, which would lead to a degradation of the temporal shape of the pulse and its duration under high-power configuration. Nevertheless, we observe once again that such a thickness is only relevant for this test case, as the resulting $\Delta t$ is outside the working regime of the ReMPI scheme.

Lastly, from a fabrication perspective, producing a fused-silica delay mask with a thickness in the 50-100 $\upmu$m range is practically achievable. The required geometry can be produced either by directly machining the different mask sections or by using a thicker baseline slab with additional annular sections to introduce the relative delay, as proposed in \cite{Marasciulli2021}. In both cases, standard AR coatings are available that can provide reflectivities of approximately $0.5\%$, while high-performance V-coatings can reduce this percentage to approximately $0.1\%$, compared with $\sim3.4\%$ for uncoated fused silica. Alternative optical glasses, such as AS87 Neo or AF32 (Schott), may also be considered, as they are more readily available in such thicknesses. In principle, pulse-to-pulse delays in the few-femtosecond to low-tens-of-femtoseconds range could be achieved by reducing the thickness difference between the mask sections; however, the availability of suitable optical materials that can be reliably fabricated at such thicknesses and safely employed in experiments makes their practical realization challenging.

\section{Conclusion}

The results presented in this work demonstrate the feasibility of a two-section delay mask for generating an ultrashort two-pulse train, providing a basis for its potential application to the ReMPI scheme. For this purpose, the transverse fluence distribution of the laser was characterized at a \mbox{100 TW} power level using EBT4 radiochromic film, enabling measurement of the beam profile without attenuation or downsizing. The obtained distribution was fitted and used as input for simulations as the basis for the mask design.

The resulting mask geometry was designed to provide comparable peak intensities of the temporally delayed beam components in the focal plane, accounting for diffraction and the different effective focusing conditions of the beam sections. Spatial characterization of the corresponding beam components was subsequently performed using equivalent hard-aperture masks at low laser power, below the threshold of the CCD sensor. The measured focal spots showed good agreement with the simulations, although a notable difference in the peak intensities was observed. This discrepancy is attributed primarily to the different laser operating conditions and may also be influenced by changes in the transverse fluence distribution and wavefront aberrations between the high- and low-power configurations.

Finally, the temporal separation introduced by the transparent delay mask was experimentally characterized using a second-order autocorrelator. The measured pulse-to-pulse delay was in close agreement with the theoretical value based on the actual mask thickness and its optical properties, while no significant pulse broadening due to propagation through the fused-silica mask was observed. These results demonstrate the capability of the proposed approach to generate temporally delayed ultrashort pulses while preserving their temporal characteristics.

\section*{Acknowledgements}
The authors acknowledge funding from the European Union's Horizon Europe research and innovation program and the UKRI guarantee funds through the MSCA ''EuPRAXIA Doctoral Network''-EuPRAXIA-DN, under Grant Agreement No. 101073480, and additional financial support from the following research programs: EU Horizon 2020 Research and Innovation Program EuPRAXIA Preparatory Phase, under Grant Agreement No. 101079773, and EU Horizon IFAST, under Grant Agreement No. 101004730. This research has been co-funded by the European Union-NextGeneration EU through the Italian PNRR MUR projects ''Integrated infrastructure initiative in Photonic and Quantum Sciences''-I-PHOQS (IR0000016, ID D2B8D520); ''EuPRAXIA Advanced Photon Sources''-EuAPS (IR0000030, CUP I93C21000160006); ''Tuscany Health Ecosystem-Spoke 1: Advanced Radiotherapies and Diagnostics in Oncology'' (ECS00000017); and the Italian INFN CSN5-funded project FRIDA.

\printcredits

\section*{Declaration of Competing Interest}
The authors declare that they have no known competing financial interests or personal relationships that could have appeared to influence the work reported in this paper.

\section*{Data Availability}
Data will be made available upon reasonable request.





\end{document}